\documentclass{ws-procs9x6-cpt19}
\usepackage{accents}
\usepackage{color}
\usepackage{xcolor}

\def\al{\alpha}
\def\be{\beta}

\def\ta{\tau}

\def\lsim{\mathrel{\rlap{\lower4pt\hbox{\hskip1pt$\sim$}}
    \raise1pt\hbox{$<$}}}
\def\gsim{\mathrel{\rlap{\lower4pt\hbox{\hskip1pt$\sim$}}
    \raise1pt\hbox{$>$}}}

\newcommand{\beq}{\begin{eqnarray}}
\newcommand{\eeq}{\end{eqnarray}}

\def\ismean{\accentset{\circ}{a}^{(3)}} 
\def\ismecn{\accentset{\circ}{c}^{(4)}} 
\def\ismetn{\accentset{\circ}{a}^{(5)}}
\def\ismegn{\accentset{\circ}{c}^{(6)}}

\begin{document}

\title{Test of Lorentz Violation with Astrophysical Neutrino Flavor in IceCube}

\author{
  Teppei Katori$^1$,
  Carlos~A.~Arg\"{u}elles$^2$,
  Kareem Farrag$^{1,3}$, and
  Shivesh Mandalia$^1$
}

\address{
  $^1$Queen Mary University of London, E1 4NS, UK\\
  $^2$Massachusetts Institute of Technology, Cambridge, MA 02139, USA \\
  $^3$University of Southampton, Southampton, SO17 1BJ, UK
}

\author{On behalf of the IceCube Collaboration}

\begin{abstract}
Astrophysical high-energy neutrinos observed by IceCube are sensitive to small effects in a vacuum such as those motivated from quantum gravity theories. Here, we discuss the potential sensitivity of Lorentz violation from the diffuse astrophysical neutrino data in IceCube. The estimated sensitivity reaches the Planck scale physics motivated region, providing IceCube with real discovery potential of Lorentz violation.
\end{abstract}

\bodymatter


\section{Neutrino interferometry}
Neutrinos make a natural interferometric system. Their production and detection occur in their flavor eigenstates, but they propagate in their Hamiltonian eigenstates. Thus, tiny disturbances during their propagation, for example tiny couplings with quantum gravity motivated physics in the vacuum, could end up with unexpected flavour composition at detection~\cite{Katori:2012pe}.

Astrophysical neutrinos propagate $\sim\mathcal{O}(100)$~Mpc, resulting in these neutrinos to become incoherent at the detection and so the phase information is washed out. However, an incoherent neutrino mixing is caused by an effective Hamiltonian which may include a potential Lorentz violating couplings of neutrinos~\cite{Kostelecky:2011gq}.
Thus, information of Lorentz violation is imprinted on the flavour composition of astrophysical neutrinos measured at the Earth. The goal of this analysis is to find nonzero SME coefficients
from the flavor data of astrophysical neutrinos in IceCube.


\section{Astrophysical neutrino flavor new physics sensitivity}
\begin{figure}
\begin{center}
\includegraphics[width=4in]{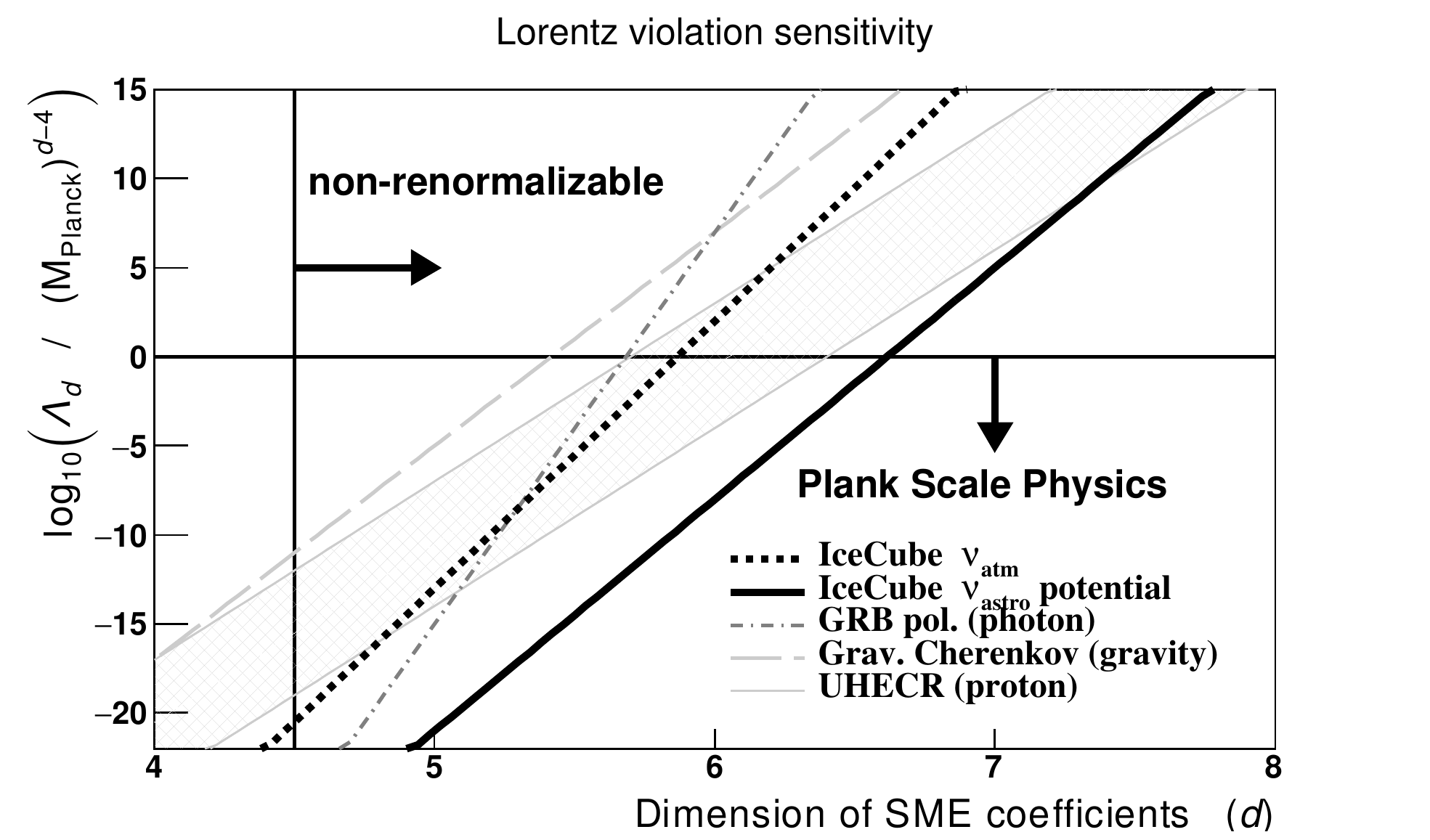}
\end{center}
\caption{Maximum sensitivity comparison for different Lorentz violation tests. Here, the x-axis is the dimension of operators, and y-axis is the new physics scale normalized by powers of the Planck mass ($\Lambda_d/M_{Planck}^{d-4}$). The solid line is the expected sensitivity of astrophysical neutrinos in IceCube, and the dashed solid line is for atmospheric neutrino limits from IceCube~\cite{Aartsen:2017ibm}. The dashed gray line is the limit from vacuum gravitational Cherenkov radiation~\cite{Kostelecky:2015dpa}, the dashed-dotted gray line is from the gamma ray burst (GRB) polarization analysis~\cite{Kostelecky:2013rv}, and the gray band is estimated from ultra-high-energy cosmic ray (UHECR) spectrum~\cite{Maccione:2009ju}.}
\label{fig:sensitivity}
\end{figure}

Fig.~\ref{fig:sensitivity} shows a naive estimation of maximum sensitivity of different methods to look for Lorentz violation. Our focus is to perform the most sensitive test of Lorentz violation, hopefully reaching into the discovery region. Here, the x-axis is the dimension of effective operators (denoted $d$), and y-axis is the order of the operator scale ($\Lambda_d$) normalized to the Planck mass ($M_{Planck}\sim 1.2\times 10^{19}$~GeV). For example, ``0'' on the y-axis for a dimension-six operator ($d=6$) corresponds to $\sim 10^{-38}$~GeV$^{-2}$.
Note, such definition makes sense only for non-renormalizable operators (operators with dimension greater than 4), which is traditionally used to look for new physics. 
A system which reaches a smaller scale in this figure has a better sensitivity to new physics. As a general trend, high energy sources are more sensitive to higher dimension operators. Neutrinos get extra sensitivity due to their interferometric nature. The solid line describes the naive sensitivity of astrophysical neutrino flavor physics on Lorentz violation, whose sensitivity exceeds any known sectors at dimension five, six, and seven operators. Furthermore, their sensitivities reach the expected region of Planck scale physics for dimension five and six operators, giving this analysis a real potential to discover quantum gravity.

\section{Astrophysical Neutrino Flavor Triangle}
\begin{figure}
\begin{center}
  \vspace{-15pt}
  \includegraphics[width=3in]{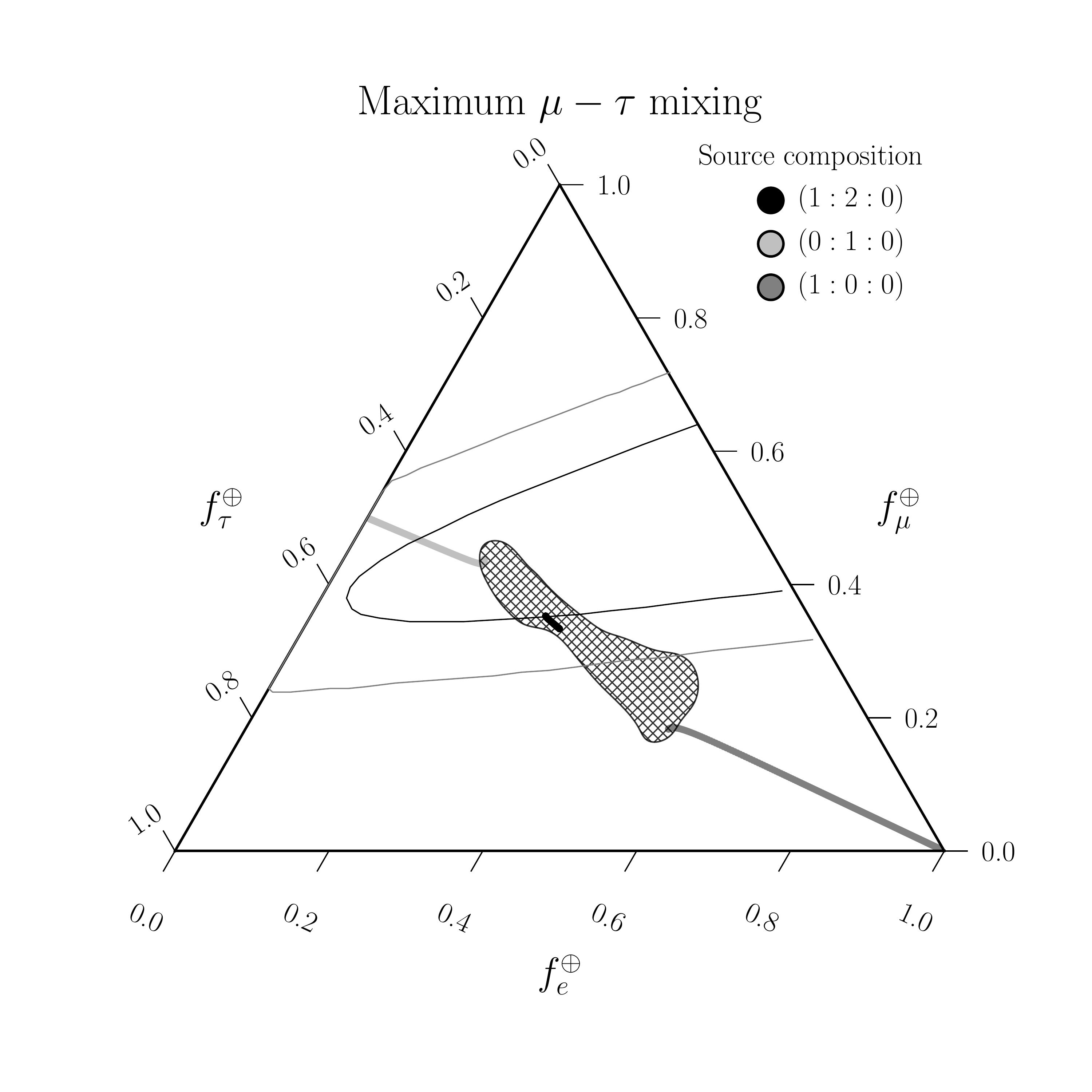}
  \vspace{-25pt}
\end{center}
\caption{Flavor triangle ternary diagram for astrophysical neutrinos. Here, the hatched region includes all possible scenarios of observable flavor ratio on the Earth by assuming the production flavor ratio $(x:1-x:0)$, $0<x<1$ and $\nu$SM oscillation parameters~\cite{Esteban:2018azc}. 68\% and 95\% contours are from IceCube data~\cite{Aartsen:2015knd}. Here, we show 3 scenarios with Lorentz violation. The black line assumes the production flavor ratio is $(0.33:0.66:0)$, whereas the gray line assumes $(1:0:0)$ and light gray line assumes $(0:1:0)$.}
\label{fig:triangle}
\end{figure}

The main parameter of this analysis is the fraction of observed neutrino flavors (flavor ratio) of astrophysical neutrinos ($\nu_e:\nu_\mu:\nu_\ta$) displayed in the flavor triangle ternary diagram (Fig.~\ref{fig:triangle}). Each corner represents a pure flavor state of astrophysical neutrino flux; for example, the bottom right corner is pure electron composition, in the ratio $\nu_e$ ($1:0:0$). Since we want to measure an unexpected flavor ratio due to Lorentz violation, which is represented a point in this diagram, we must {\it a priori} know the flavor ratio without new physics. To predict the flavor ratio on Earth, first, we need to know the flavor ratio at the production. This is expected somewhere between $\nu_e$ dominant and $\nu_\mu$ dominant scenarios (toward the right axis of the triangle). Thus, a generic model such that ($x:1-x:0$) with $0<x<1$ can describe all possible production models. Secondly, without assuming any new physics, neutrinos mix by the neutrino masses. We use the neutrino mass parameters from global oscillation data fit within the neutrino Standard Model ($\nu$SM)~\cite{Esteban:2018azc}. The combination of these two makes the hatched region. Since the central region of the triangle has the highest phase space density, any production models with reasonable assumptions (including three flavors, unitarity, etc.) will end up in the central area in this diagram~\cite{Arguelles:2015dca}. 
We show 68\% and 95\% contours of the flavor ratio measurement from an analysis in IceCube~\cite{Aartsen:2015knd}. Most of hatched region is contained in this contour, meaning current analyses don't have enough power to distinguish different production scenarios of astrophysical neutrinos within $\nu$SM.

Now, we introduce Lorentz violation. As an example, we introduce an isotropic SME coefficient of dimension-six operator with maximum $\nu_\mu-\nu_\tau$ mixing. We assume all astrophysical neutrinos follow an $\sim E^{-2}$ spectrum within [60~TeV,~10~PeV] and the scale of isotropic dimension-six SME coefficient $\ismegn$ are varied from $10^{-52}$~GeV$^{-2}$ (i.e., very small) to $10^{-42}$~GeV$^{-2}$ (to very large). We also assume three standard astrophysical neutrino production models; $\nu_e$ dominant ($(1:0:0)$ (gray), $\nu_\mu$ dominant ($(0:1:0)$ (light gray), and pion decay models $(0.33:0.66:0)$ (black). When the scale of the SME coefficient is too small, it would cause no effects on the observable flavor ratio and all 3 scenarios end up in the hatched region. Once the value starts to increase, some of them start to leave the hatched region and the contour. If this is the case, we would observe a large deviation of astrophysical neutrino flavor ratio from standard scenarios and hence could discover nonzero Lorentz violation. Clearly, experiments need to strive to shrink this contour. Improving knowledge of oscillation parameters will help to shrink the hatched region. In this example, we expect to find nonzero SME coefficients if it causes $\mu-\tau$ mixing under the assumption of a high $\nu_e$ component at the production. Note, the formalism used in this study is applicable to look for other types of new physics (for example~\cite{Rasmussen:2017ert,Klop:2017dim,Bustamante:2018mzu,Farzan:2018pnk}).


\vspace{-25pt}
\begin{thebibliography}{10}
\bibitem{Katori:2012pe}
T.~Katori, 
  {\em Mod. Phys. Lett.}, vol.~A27, p.~1230024, 2012.
  
\bibitem{Kostelecky:2011gq}
A.~Kosteleck\'{y} and M.~Mewes, 
  {\em Phys. Rev.}, vol.~D85, p.~096005, 2012.
  
\bibitem{Aartsen:2017ibm}
M.~G. Aartsen {\em et~al.}, 
  {\em Nature Phys.}, vol.~14, no.~9, pp.~961--966, 2018.

\bibitem{Kostelecky:2015dpa}
V.~A. Kosteleck\'{y} and J.~D. Tasson, 
  {\em Phys. Lett.}, vol.~B749, pp.~551--559, 2015.
  
\bibitem{Kostelecky:2013rv}
V.~A. Kosteleck\'{y} and M.~Mewes, 
  {\em Phys. Rev. Lett.}, vol.~110, p.~201601, 2013.
  
\bibitem{Maccione:2009ju}
L.~Maccione {\it et al.}, 
  {\em JCAP}, vol.~0904, p.~022, 2009.
  
\bibitem{Esteban:2018azc}
I.~Esteban {\it et al.}, 
  {\em JHEP}, vol.~01, p.~106, 2019.
  
\bibitem{Arguelles:2015dca}
C.~Arg\"{u}elles, T.~Katori, J.~Salvado, 
  {\em Phys. Rev. Lett.},~115,~161303, 2015.
  
\bibitem{Aartsen:2015knd}
M.~G. Aartsen {\em et~al.},
  {\em Astrophys. J.}, vol.~809, no.~1, p.~98, 2015.
  
\bibitem{Rasmussen:2017ert}
R.~W. Rasmussen {\it et al.}, 
{\em Phys. Rev.}, vol.~D96, no.~8, p.~083018, 2017.

  \bibitem{Klop:2017dim} 
  N.~Klop and S.~Ando,
  Phys.\ Rev.\ D {\bf 97}, no. 6, 063006 (2018)
  
\bibitem{Bustamante:2018mzu}
M.~Bustamante and S.~K. Agarwalla, 
{\em Phys. Rev. Lett.}, 122, 061103, 2019.

\bibitem{Farzan:2018pnk} 
  Y.~Farzan and S.~Palomares-Ruiz,
  Phys.\ Rev.\ D {\bf 99}, no. 5, 051702 (2019)
  
\end{thebibliography}
\end{document}